\documentclass[
  reprint,
  superscriptaddress,
  frontmatterverbose,
  amsmath,amssymb,
  aps,
  prfluids
]{revtex4-2}

\usepackage[normalem]{ulem}
\usepackage{xcolor}
\usepackage{dcolumn}
\usepackage{bm}
\usepackage[caption=false]{subfig}
\usepackage{float}  
\usepackage{hyperref} 
\usepackage{graphicx}
\usepackage{epstopdf}
\hypersetup{
    colorlinks=true,     
    linkcolor=blue,      
    filecolor=blue,      
    urlcolor=black,         
    citecolor=blue         
}
\usepackage[mathlines]{lineno}
\usepackage[none]{hyphenat}

\begin{document}

\preprint{}

\title{Nonlocal flow sampling enables vortex trapping of heavy particles}

\author{Sachin Kulkarni}
\thanks{These authors contributed equally to this work.}
\affiliation{Department of Mathematical and Computational Sciences, National Institute of Technology Karnataka, Surathkal 575025, India}

\author{Sumithra R. Yerasi}
\thanks{These authors contributed equally to this work.}
\affiliation{Aix Marseille Univ, CNRS, Centrale Med, IRPHE, Marseille, France}

\author{Vishwanath Kadaba Puttanna}
\affiliation{Department of Mathematical and Computational Sciences, National Institute of Technology Karnataka, Surathkal 575025, India}

\author{Dario Vincenzi}
\affiliation{Universit\'e C\^ote d’Azur, CNRS, LJAD, 06100 Nice, France}

\author{S. Ravichandran}
\affiliation{Center for Climate Studies, Indian Institute of Technology Bombay, Mumbai 400076, India}

\author{KVS Chaithanya}
\thanks{Email address for correspondence: chaithanyakvs@iiti.ac.in}
\affiliation{Department of Chemical Engineering, \\ \mbox{Indian Institute of Technology Indore,}
Indore, \mbox{Madhya Pradesh 453552}, India}

\begin{abstract}

Most analyses of inertial particle motion in vortical flows rely on the point-particle approximation, in which the fluid velocity is assumed to be linear at the scale of the particle, and for heavy particles inertia typically leads to centrifugal expulsion from vortex cores. Here, we show that a spatially extended particle, modeled as a rigid symmetric dumbbell of two identical inertial point particles connected by a massless rod that samples the flow at two points, can converge to a vortex-centered spinning state. We study the dynamics of this inertial dumbbell in a steady two-dimensional Lamb--Oseen vortex and identify three qualitatively distinct long-time behaviors controlled by the Stokes number. In the weak-inertia limit, the motion remains bounded and traces spirographic-like trajectories around the vortex center, while at sufficiently large inertia centrifugal effects dominate and trajectories spiral outward, approaching inertial point-particle behavior. Between these limits, the dumbbell can reach a trapped spinning state in which the center-of-mass converges to the vortex center and spins steadily, with accessibility determined by the initial conditions. Basin-of-attraction maps and ensemble statistics reveal a non-monotonic dependence of the accessibility of the spinning state on inertia, with basins of finite measure occurring only over an intermediate range of Stokes numbers. Linear stability is governed by the logarithmic slope of the vortex angular-velocity profile, and for the Lamb--Oseen vortex the spinning state is stable for all Stokes numbers. These results highlight how nonlocal flow sampling by spatially extended inertial particles can fundamentally alter transport and long-time behavior in vortical flows.

\end{abstract}

\maketitle

\section{Introduction}

Transport and dispersal of particles in turbulent flows are central to a wide range of atmospheric, oceanic, and industrial processes, motivating extensive studies of particle motion and organization in turbulence (see, e.g., \cite{mathai2020bubbly,brandt2022particle} and references therein). Within turbulent flows, vortices represent a key class of flow structures affecting particle dynamics and provide a natural setting for studying particle transport in a simplified framework. Particle dynamics in vortical flows have predominantly been studied within the point-particle framework~\cite{rdg17}. However, many particles encountered in natural and industrial settings have a finite spatial extent and different parts of the particle experience different velocities. This nonlocal flow sampling generically couples translational and rotational dynamics and can substantially affect the particle's motion, motivating models beyond the point-particle approximation. Solving the dynamics of an object extended in space is a highly complex problem. Therefore, to gain insight into the interplay between nonlocal flow sampling and inertia, in this study,  we investigate a simple model of an extended particle: an inertial rigid dumbbell, consisting of two inertial point particles connected by a rigid, massless rod, that samples the flow at two distinct points, in a steady Lamb–Oseen vortex.

The dynamics of point particles in vortical flows are relatively well understood. In the absence of inertia, they behave as tracers and follow streamlines; when inertia is present, heavy particles move across streamlines and are expelled from vortex cores~\cite{rm97, gl04, ravichandran2015caustics, ravichandran2022waltz}. At sufficiently large distances from the vortex center, their trajectories asymptotically approach Fermat spirals~\cite{ravichandran2022waltz}. Trapping in the vicinity of the vortex core is observed when the vortex is not axisymmetric~\cite{nr25} or in the presence of multiple vortices \cite{angilella2010dust, ravichandran2014attracting, kapoor2024trapping,knpr25}.
More generally, in turbulent flows, heavy point particles preferentially accumulate in regions of low vorticity or high strain, a phenomenon commonly referred to as preferential concentration or inertial bias~\cite{pw16,bgm24}. 

Particles with finite spatial extent sample the flow over a finite length and are therefore affected by velocity differences. The resulting dynamics have been investigated in nonlinear laminar flows for a range of elongated particles, including rigid fibers~\cite{lopez2017inertial, com20, schreder25}, flexible fibers~\cite{duroure2019}, and dumbbells~\cite{dlmv02,pg03,piva2008rigid, yerasi2022spirographic}, as well as for discs~\cite{ibg23} and deformable spheres~\cite{famg24}. Despite these advances, the motion of an extended body remains a challenging problem even for the simplest flows, and several questions are still open. Here we focus on the effect of inertia on spatially extended particles in vortical flows. We study how the dynamics deviates from that of inertial point particles, given that the particle can sample the nonlinearity of the carrier velocity field.

The rigid dumbbell model, consisting of two point particles (beads) linked by a rigid, massless rod, retains much of the simplicity of point-particle dynamics while introducing a finite length scale that enables sampling of velocity differences in the surrounding flow. When bead inertia is included, dumbbell dynamics have been studied primarily in the context of gravity-driven motion in a quiescent fluid~\cite{cm16} or in cellular flows~\cite{dlmv02,pg03,piva2008rigid}, where periodic and chaotic trajectories have been reported. In vortical flows, dumbbell dynamics have so far been examined mainly in the inertia-free limit, where bounded, space-filling, spirographic-like trajectories are observed~\cite{yerasi2022spirographic}. These trajectories differ qualitatively from the tracer-like motion observed for point particles.

In this study, we examine a rigid inertial dumbbell in a steady Lamb–Oseen vortex. Starting from the equations of motion for the two beads, we write coupled evolution equations for the dumbbell’s center-of-mass position and orientation. Inertia is quantified by the Stokes number, defined as the ratio of particle and flow time scales. For small Stokes numbers, the motion reduces to the spirographic-like trajectories reported for inertia-free dumbbells, whereas at large Stokes numbers centrifugal expulsion dominates, as in the inertial point-particle case. Between these limits, we identify a distinct regime in which the dumbbell can become trapped at the vortex center and spin steadily, which we term the \textit{spinning} state.

This manuscript is organized as follows. In Sec.~\ref{Sec:Mathematical_Modelling}, we introduce the inertial dumbbell model and write the coupled translational and rotational equations of motion in a Lamb--Oseen vortex. In Sec.~\ref{sect:results}, we analyze the dynamics through center-of-mass trajectories and temporal evolution, identify distinct dynamical regimes including the vortex-centered spinning state, and characterize its basin of attraction. The linear stability of the spinning state is then examined analytically. Finally, Sec.~\ref{sec:Conclusions} summarizes the main findings.

\begin{figure}[t]
\centering
 \includegraphics[width=0.6\linewidth]{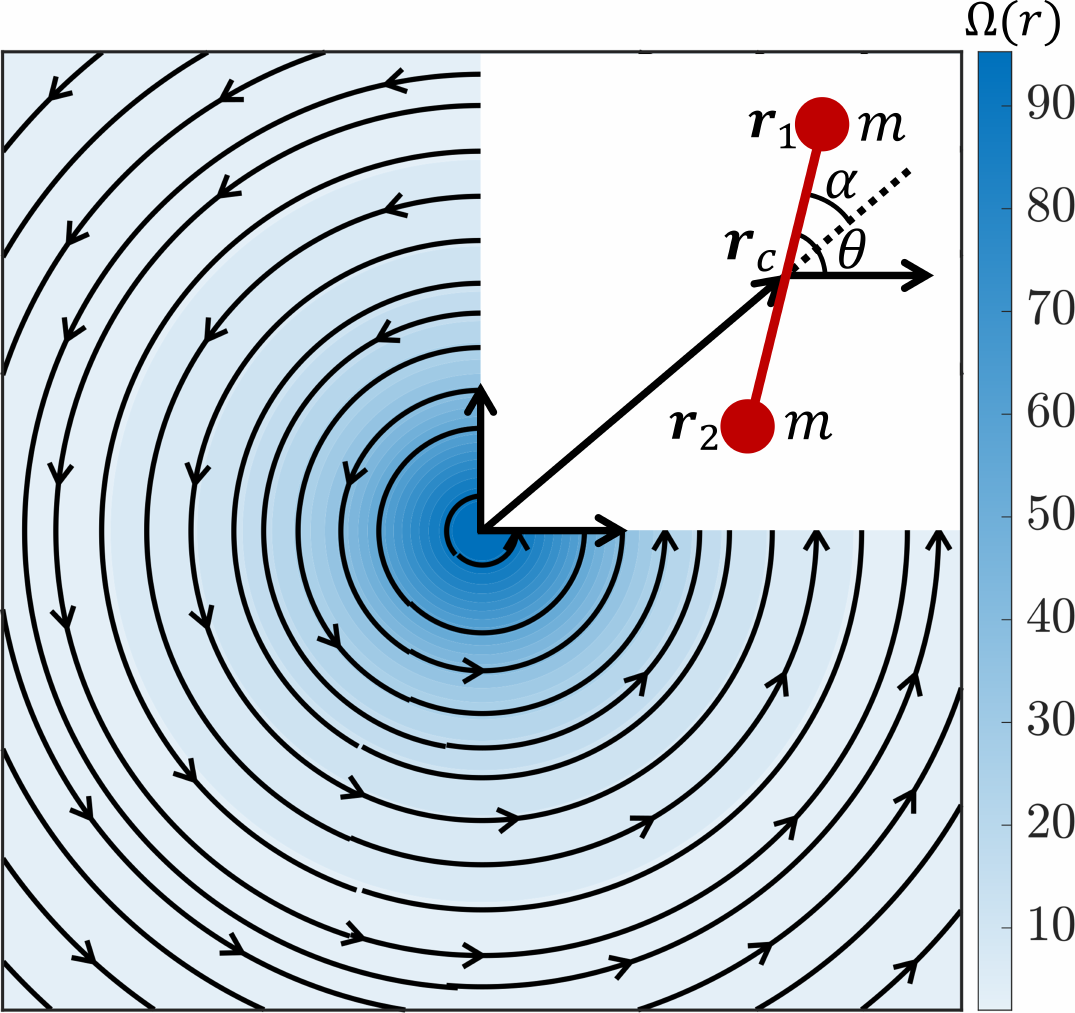}
  \captionsetup{justification=raggedright}
\caption{\label{fig:schematic} Schematic of a rigid dumbbell of length $\ell$ in a Lamb--Oseen vortex. The color map indicates the fluid angular velocity $\Omega(r)$. The inset shows the coordinate system and notation: $\bm{r}_1$ and $\bm{r}_2$ denote the bead positions, and $\bm{r}_c$ denotes the center-of-mass position. The dumbbell orientation is given by the angle $\theta$ measured from the $x$-axis, and $\alpha$ is the angle between the rod and $\bm{r}_c$.}
\end{figure}

\section{Mathematical Modeling}
\label{Sec:Mathematical_Modelling}

We consider the motion of a rigid dumbbell composed of two identical beads of radius $R_b$ and mass $m$, connected by a rigid rod of fixed length $\ell$, evolving in a steady Lamb--Oseen vortex (Fig.~\ref{fig:schematic}). The beads are assumed to be heavy, i.e., their density exceeds that of the surrounding fluid. The rod is treated purely as a geometric constraint: it enforces a constant separation between the beads but does not itself interact with the fluid. We assume $\ell \gg R_b$ and neglect hydrodynamic interactions between the beads.

In the heavy-particle limit, the motion of each bead is governed by a balance between inertia, Stokes drag, and the tension force exerted by the rigid rod; added-mass effects and the Boussinesq--Basset history term are neglected. The Stokes drag force on bead $i$ is
\begin{equation}
\bm{F}_i = -\zeta \bigl[\dot{\bm{r}}_i - \bm{u}(\bm{r}_i)\bigr],
\end{equation}
where $\bm{r}_i\equiv\bm{r}_i(t)$ denotes the position of the $i$-th bead ($i=1,2$) at time $t$, $\zeta$ is the drag coefficient, $\bm{u}(\bm{r}_i)$ is the background fluid velocity evaluated at the bead position, and an overdot denotes differentiation with respect to time. The equation of motion for bead $i$ is
\begin{equation}\label{eqn:inertiaeqnofmotionbeads}
m \ddot{\bm{r}}_i = -\zeta \bigl[\dot{\bm{r}}_i - \bm{u}(\bm{r}_i)\bigr] + \bm{\tau}_i,
\qquad i = 1,2,
\end{equation}
where $\bm{\tau}_i$ is the tension force exerted by the rod on the $i$-th bead.

\begin{figure*}[t]
\centering
 \includegraphics[width=\linewidth]{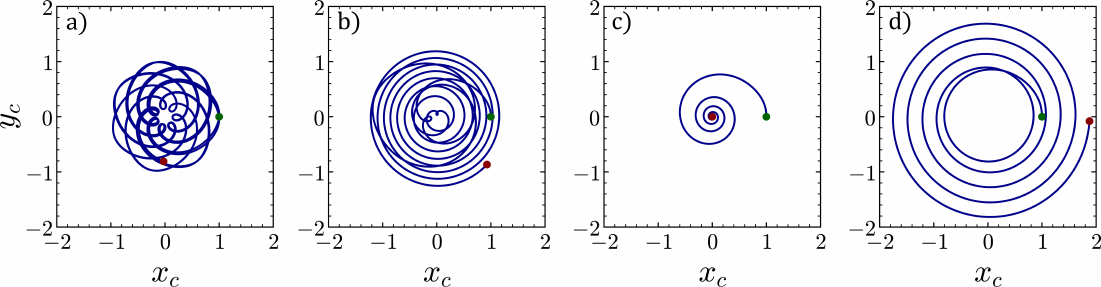}
\caption{\label{fig:trajectories} Trajectories of the center-of-mass of the dumbbell in the $(x_c,y_c)$ plane for
different Stokes numbers:
(a) $\mathrm{St}=10^{-5}$,
(b) $\mathrm{St}=10^{-2}$,
(c) $\mathrm{St}=5\times10^{-2}$, and 
(d) $\mathrm{St}=10^{-1}$.
The green marker indicates the initial position of the center-of-mass at
$(x_c(0),y_c(0))=(1,0)$, and the red marker denotes the position at time
$t=50$. At $t=0$, the orientation is $\theta(0)=0$, and the initial translational
and angular velocities are zero, $v_{cx}(0)=v_{cy}(0)=0$ and $\omega(0)=0$.
}
\end{figure*}

Equation~\eqref{eqn:inertiaeqnofmotionbeads} can be recast in terms of the center-of-mass and rotational degrees of freedom of the dumbbell. We define the center-of-mass position and velocity as
$\bm{r}_c = (\bm{r}_1 + \bm{r}_2)/2$ and $\bm{v}_c = (\bm{v}_1 + \bm{v}_2)/2$, respectively, where $\bm{v}_i=\dot{\bm{r}}_i$.
The rod orientation is described by $\theta(t)$, measured relative to the $x$-axis (see Fig.~\ref{fig:schematic}), with angular velocity $\omega=\dot{\theta}$.
Adding Eq.~\eqref{eqn:inertiaeqnofmotionbeads} for beads 1 and 2 and using $\bm{\tau}_1 = -\bm{\tau}_2$ yields the translational equation of motion for the center-of-mass,
\begin{equation}\label{eqn:inertiacenterofmass}
	m \dot{\bm{v}}_c
	=
	-\zeta \left[
	\bm{v}_c
	-
	\frac{1}{2}\bigl(\bm{u}(\bm{r}_1) + \bm{u}(\bm{r}_2)\bigr)
	\right].
\end{equation}

The rotational dynamics follow from the angular momentum balance about the center-of-mass,
\begin{equation}\label{eqn:dLdt}
\dot{\bm{L}} = \bm{M}_1 + \bm{M}_2,
\end{equation}
where $\bm{L}$ is the angular momentum of the dumbbell about $\bm{r}_c$ and $\bm{M}_i = (\bm{r}_i-\bm{r}_c)\times \bm{F}_i$ is the torque due to the drag on bead $i$. We consider planar motion in the $x$--$y$ plane, for which $\boldsymbol{\omega}=\omega\,\bm{\hat{z}}$ and $\bm{L}=I\boldsymbol{\omega}$,
where $I = \frac{4}{5} m R_b^2 + m \frac{\ell^2}{2}$ is the moment of inertia about the center-of-mass. In the limit $R_b \ll \ell$, the inertia is dominated by the parallel-axis contribution of the beads and reduces to $I \approx m\ell^2/2$.
Writing $\bm{\ell}=\bm{r}_1-\bm{r}_2 =\ell(\cos\theta,\sin\theta)$ with $\ell=|\bm{\ell}|$ and using $\bm{\ell}\times\dot{\bm \ell}=\ell^2\omega\bm{\hat{z}}$ reduces Eq.~\eqref{eqn:dLdt} to
\begin{equation}\label{eqn:inertiaangularvel}
I \dot{\omega}
=
\dfrac{\zeta}{2}
\left[
\bm{\ell} \times \bigl(\bm{u}(\bm{r}_1)-\bm{u}(\bm{r}_2)\bigr)\cdot \bm{\hat{z}}
-
\ell^2 \omega
\right].
\end{equation}
The background flow is assumed to be a steady, two-dimensional Lamb--Oseen vortex with center at the origin and angular-velocity profile
\begin{equation}
\Omega(r) = \frac{\Gamma}{2\pi}\,\frac{1-\mathrm{e}^{-r^2/R^2}}{r^2},
\end{equation}
where $r = |\bm{r}|$. Here $\Gamma$ is the circulation and $R$ is the vortex core size. The corresponding velocity field is $\bm{u}(\bm{r}) = r\,\Omega(r)\,\bm{\hat{\phi}}$, where $\bm{\hat{\phi}}$ is the azimuthal unit vector.

Nondimensionalizing Eqs.~\eqref{eqn:inertiacenterofmass} and \eqref{eqn:inertiaangularvel} using $\ell$ as the characteristic length and $U_0= \Gamma/(2\pi\ell)$ as the characteristic velocity yields~\cite{piva2008rigid} (for simplicity, we use the same notation for the dimensionless variables as their dimensional counterparts)
\begin{subequations}
\label{eqn:ndeqns}
\begin{align}
    \dot{\bm{v}}_c &= \ddot{\bm{r}}_c
    = \frac{1}{\mathrm{St}}\left[\frac{1}{2}\bigl(\bm{u}(\bm{r}_1) + \bm{u}(\bm{r}_2)\bigr) - \bm{v}_c \right], \\
    \label{eq:ndeqns-omega}
    \dot{\omega} &= \ddot{\theta}
    = \frac{1}{2\mathrm{St}}\left[\bm{\ell} \times \bigl(\bm{u}(\bm{r}_1) - \bm{u}(\bm{r}_2)\bigr)\cdot \bm{\hat{z}} - \omega \right].
\end{align}
\end{subequations}
Here, $\mathrm{St} = m U_0 / (\zeta \ell)$ is the Stokes number, defined as the ratio of the particle inertial time scale $m/\zeta$ to the characteristic flow time scale $\ell/U_0$.

Equations~\eqref{eqn:ndeqns} constitute a system of six ordinary differential equations for the center-of-mass position $(x_c,y_c)$ and translational velocity $(v_{cx},v_{cy})$, and the rod orientation $\theta$ and angular velocity $\omega$. The system is numerically integrated using the \texttt{solve\_ivp} solver in Python with a fifth-order Runge--Kutta scheme (RK45), with relative and absolute tolerances set to $\texttt{rtol}=10^{-9}$ and $\texttt{atol}=10^{-12}$.
Based on the study of the inertia-free case \cite{yerasi2022spirographic}, we expect the system to display a sensitive dependence on both the parameters and the initial conditions. Here we focus on the effect of inertia, i.e., we examine the dynamics of the dumbbell as St is varied. Therefore,
unless otherwise specified, we set $R=0.1$, and the initial conditions are $x_c(0)=1$, $y_c(0)=0$, $\theta(0)=0$, with zero initial translational and angular velocities.

\section{Results and Discussion}
\label{sect:results}

\subsection{Center-of-Mass Trajectories}

\begin{figure*}[t]
\centering
 \includegraphics[width=\linewidth]{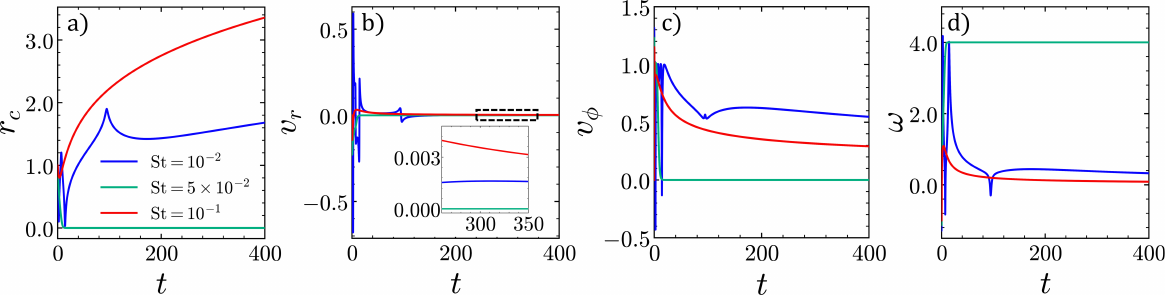}
\caption{\label{fig:time_series} Long-time evolution of a dumbbell in a Lamb--Oseen vortex for $\mathrm{St}=10^{-2}$ (blue), $\mathrm{St}=5 \times 10^{-2}$ (green), and $\mathrm{St}=10^{-1}$ (red), with the same initial conditions as in Fig.~\ref{fig:trajectories}. Panels show (a) the radial distance $r_c(t)$, (b) the radial velocity $v_r(t) = \dot{r}_c$ and (c) the azimuthal velocity $v_\phi(t) = r_c \dot{\phi}_c$ of the dumbbell center-of-mass, and (d) the angular velocity $\omega(t) = \dot{\theta}$ of the dumbbell. The inset in (b) magnifies the late-time interval corresponding to the dashed box.}
\end{figure*}

Figure~\ref{fig:trajectories} shows representative center-of-mass trajectories in the $(x_c,y_c)$ plane as the Stokes number increases. For $\mathrm{St}=10^{-5}$, the motion is bounded: the trajectory executes repeated loops that gradually fill an annular region, producing a compact spirographic pattern [Fig.~\ref{fig:trajectories}(a)], consistent with the inertia-free dumbbell dynamics~\cite{yerasi2022spirographic}. At $\mathrm{St}=10^{-2}$, the motion is initially similar, but inertia induces a net outward drift such that the trajectory becomes unbounded and spirals outward over many revolutions [Fig.~\ref{fig:trajectories}(b)]. At an intermediate value, $\mathrm{St}=5\times10^{-2}$, the center-of-mass spirals inward and settles at the vortex center, while the dumbbell continues to spin about its center-of-mass; we refer to this long-time behavior as the {\it spinning} state [Fig.~\ref{fig:trajectories}(c)]. Increasing the inertia further to $\mathrm{St}=10^{-1}$ restores an unbounded trajectory, with the center-of-mass spiraling away while continuing to orbit the vortex center [Fig.~\ref{fig:trajectories}(d)].

Thus, as the Stokes number is increased, the long-time dynamics change in a non-monotonic manner: spirographic motion occurs in the negligible-inertia limit, while at finite inertia both outward spiraling and convergence to a vortex-centered spinning state are observed, with outward spiraling dominating again at larger inertia and approaching inertial point-particle behavior.

\subsection{Temporal Dynamics of the Dumbbell}

To understand the dynamics that correspond to the trajectories described in the previous Section, it is useful to decompose the motion of the center-of-mass of the dumbbell into its radial and tangential components. We write the center-of-mass position in polar form as $\bm r_c=r_c(\cos\phi_c, \sin\phi_c)$. The corresponding velocity components are $v_r=\dot r_c$ and $v_\phi=r_c\dot\phi_c$. Thus, in Fig.~\ref{fig:time_series}, we show the time evolution of the center-of-mass radial position, radial and azimuthal velocities, and the dumbbell's angular velocity. The Stokes numbers are the same as in Fig.~\ref{fig:trajectories}, excluding \(\mathrm{St}=10^{-5}\). For \(\mathrm{St}=10^{-5}\), all variables simply exhibit fully periodic time series, consistent with the closed spirographic trajectories in Fig.~\ref{fig:trajectories}(a).

The center-of-mass radial position $r_c(t)$ allows us to distinguish the spinning state from outward spiraling motions [see Fig.~\ref{fig:time_series}(a)].
For both $\mathrm{St}=10^{-2}$
and $\mathrm{St}=10^{-1}$, $r_c(t)$ grows steadily at long times, consistent with the outward spiraling motions in Figs.~\ref{fig:trajectories}(b),(d). However, when inertia is lower, a pronounced transient is present with oscillations and an initial close approach to the vortex center. When inertia is more prominent the non-monotonic transient is much shorter. The behavior is very different for the intermediate Stokes number $\mathrm{St}=5\times10^{-2}$. In this case, $r_c(t)$ decays rapidly to zero and remains there indefinitely, indicating trapping of the center-of-mass at the vortex center [the spinning state in Fig.~\ref{fig:trajectories}(c)]. 

Figure~\ref{fig:time_series}(b) shows the corresponding time evolution of the radial velocity $v_r(t)$, which underlies the trends in $r_c(t)$ shown in Fig.~\ref{fig:time_series}(a). For $\mathrm{St}=10^{-2}$, $v_r(t)$ exhibits early-time oscillations with decaying amplitude and approaches a small positive value at late times [see inset Fig.~\ref{fig:time_series}(b)].
This behavior is consistent with the eventual outward drift of the center-of-mass and the unbounded trajectory in Fig.~\ref{fig:trajectories}(b). For $\mathrm{St}=5\times10^{-2}$, $v_r(t)$ decays to zero, which confirms the trapping of the dumbbell at the vortex center for this value of St. Finally, for $\mathrm{St}=10^{-1}$, $v_r(t)$ shows a short transient before approaching again a finite positive value; a sustained outward spiraling motion is indeed observed at large inertia.

\begin{figure*}[t]
\centering
\includegraphics[width=\linewidth]{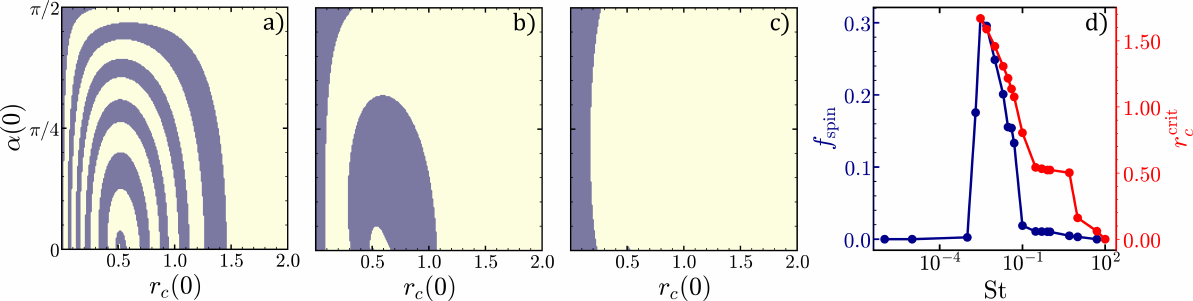}
\caption{ \label{fig:basins} Basins of attraction of the spinning state at the vortex center in the plane of initial conditions $(r_c(0),\alpha(0))$ for different Stokes numbers: (a) $\mathrm{St}=10^{-2}$, (b) $\mathrm{St}=5\times 10^{-2}$, and (c) $\mathrm{St}=1$. Purple and yellow regions indicate, respectively, initial conditions that lead to the trapped spinning state at the origin and the outward-spiraling state. (d) Fraction of initial conditions leading to the spinning state, $f_{\mathrm{spin}}$ (blue, left axis), and the corresponding critical initial radius $r_c^{\mathrm{crit}}$ (red, right axis, see text for definition) as functions of the Stokes number.}
\end{figure*}

 The time evolution of the azimuthal velocity of the center-of-mass, $v_\phi(t)$, is shown in Fig.~\ref{fig:time_series}(c). For $\mathrm{St}=5\times10^{-2}$, $v_\phi(t)$ obviously decays to zero, since the center-of-mass becomes stationary at the vortex center. For $\mathrm{St}=10^{-2}$ and $\mathrm{St}=10^{-1}$, $v_\phi(t)$ approaches a finite nonzero value, indicating continued orbital motion as the center-of-mass spirals outward. Indeed, at late times, all outward-spiraling trajectories satisfy \( r_c v_\phi \to 1 \) (see Fig.~\ref{fig:appendixB} in Appendix~\ref{sec:AppendixB}). This fact can be understood by noting that, in the far-field limit where \( r_c \gg R\) and \(r_c \gg \ell \), the center-of-mass acceleration is negligible, see Figs.~\ref{fig:time_series}(b),(c), and therefore Eq.~\eqref{eqn:ndeqns} implies \( \bm{v}_c \simeq [\bm{u}(\bm{r}_1)+\bm{u}(\bm{r}_2)]/2 \).
 Figures~\ref{fig:time_series}(b),(c) also indicate that $v_\phi \gg v_r$ at long times.
 Moreover,  in this regime the Lamb--Oseen vortex reduces to \( \bm{u}(\bm{r}) \simeq r^{-1}\hat{\boldsymbol{\phi}} \), and the flow varies weakly across the dumbbell.
 Consequently, $v_\phi\simeq v_c \simeq (1/r_1 + 1/r_2)/2\simeq 1/r_c$, and hence \( r_c v_\phi = r_c^2\dot{\phi}_c\simeq 1 \).
We conclude that, in the outward spiraling regimes, the center-of-mass displays, at long times, the same Fermat-spiral behavior as an inertial point particle~\cite{ravichandran2022waltz}.

Figure~\ref{fig:time_series}(d) shows the time evolution of the dumbbell angular velocity $\omega(t)$. In all cases, $\omega(t)$ becomes constant at late times, with a transient behavior that depends on the Stokes number. For $\mathrm{St} = 10^{-2}$, $\omega(t)$ exhibits an initial overshoot followed by damped oscillations before relaxing to its asymptotic value, whereas at $\mathrm{St} = 10^{-1}$ it increases slightly above its late-time value during the transient and then relaxes monotonically. For $\mathrm{St} = 5\times 10^{-2}$, in the case of spinning, $\omega(t)$ approaches a steady nonzero value without oscillations. This steady-state angular velocity of the spinning state is independent of $\mathrm{St}$ and initial conditions, as follows directly from Eq.~\eqref{eqn:ndeqns}. Setting $\dot{\omega}=0$ and $r_c=0$ yields \( \omega=\Omega(r_1)=\Omega(r_2)=\Omega(1/2)\), corresponding to the fluid angular velocity evaluated at the bead locations ($r_1=r_2=1/2$). For $R=0.1$, this gives $\Omega(1/2)\simeq 4$, which corresponds to the long-time plateau observed in Fig.~\ref{fig:time_series}(d).

\subsection{Basins of Attraction}

Next, we examine the basins of attraction of the two asymptotic states identified above: a spinning state, in which the center-of-mass is trapped at the vortex center and the dumbbell rotates steadily, and an outward-spiraling state.

Figure~\ref{fig:basins} shows these basins of attraction in the plane of initial conditions spanned by the radial distance of the center-of-mass from the vortex center,
$r_c(0)=|\bm{r}_c(0)|$, and the orientation of the dumbbell relative to the radial direction at the center-of-mass, $\alpha(0)$ (see Fig.~\ref{fig:schematic}). Purple regions indicate initial conditions that  lead to the spinning state, while yellow regions correspond to trajectories that spiral outward at late times. For each Stokes number, we initialize a uniform $200\times200$ grid in $(r_c(0),\alpha(0))$, corresponding to $40\,000$ distinct initial conditions. Exploiting the fore-aft symmetry of the dumbbell, it is sufficient to consider $\alpha(0)\in[0,\pi/2]$. We also restrict the initial radii to $0\le r_c(0)\le r_{\max}$ with $r_{\max}=20 R = 2$, thereby focusing on the region where vortex-induced effects are most pronounced. Each trajectory is integrated forward in time to its asymptotic state and classified according to its long-time behavior.

At low inertia ($\mathrm{St}=10^{-2}$), Fig.~\ref{fig:basins}(a) shows that the basin boundary is sensitive to both the initial radial position $r_c(0)$ and the initial orientation $\alpha(0)$. For sufficiently small $r_c(0)$, the spinning state is observed for all $\alpha(0)$, defining a near-vortex-center trapping region. The radial extent of this region increases as $\alpha(0)\to\pi/2$, i.e., configurations initially aligned with the local azimuthal direction are trapped over a wider range of initial radii. Outside this near-core region, the basin breaks up into alternating regions corresponding to spinning and outward-spiraling trajectories, forming a nested, banded structure in the $(r_c(0),\alpha(0))$ plane. Beyond the outermost band, trajectories spiral outward for all orientations, indicating the absence of trapping at sufficiently large initial radii.

At intermediate inertia ($\mathrm{St}=5 \times 10^{-2}$), shown in Fig.~\ref{fig:basins}(b), the basin structure simplifies markedly. The near-vortex-center trapping region persists and extends to larger values of $r_c(0)$ compared to the $\mathrm{St}=10^{-2}$ case. Outside this region, the banded structure disappears, and the basin of attraction of the spinning state instead forms a single connected lobe in the $(r_c(0),\alpha(0))$ plane. For $r_c(0)\gtrsim1$, trajectories spiral outward for all orientations, indicating that trapping no longer occurs beyond this radius.

In the large-inertia limit ($\mathrm{St}=1$), shown in Fig.~\ref{fig:basins}(c), the basin of attraction of the spinning state is confined to a narrow region close to the vortex center. Only dumbbells initialized sufficiently close to the center reach the spinning state. Although this near-center region is radially thicker than at lower Stokes numbers, it occupies a much smaller fraction of the accessible initial-condition space, reflecting the reduced effectiveness of trapping at high inertia.

To quantify how the basin of attraction of the spinning state varies with inertia, we define two ensemble measures: (i) the fraction of dumbbells that converge to the trapped spinning state, $f_{\mathrm{spin}}$, and (ii) the critical initial radius $r_c^{\mathrm{crit}}$, the largest $r_c(0)$ for which there exists an orientation $\alpha(0)$ that leads to the spinning state. We compute $f_{\mathrm{spin}}$ using an area-weighted measure in
the $(r_c(0),\alpha(0))$ plane,
\begin{equation}
f_{\mathrm{spin}} \;=\;
\frac{\displaystyle
\int_{0}^{\pi/2}\!\!\int_{0}^{r_{\max}}
\mathbf{1}_{\mathrm{spin}}(r_c,\alpha)\, r_c \, \mathrm{d}r_c \, \mathrm{d}\alpha}
{\displaystyle
\int_{0}^{\pi/2}\!\!\int_{0}^{r_{\max}}
r_c \, \mathrm{d}r_c \, \mathrm{d}\alpha } \, ,
\end{equation}
where $\mathbf{1}_{\mathrm{spin}}(r_c,\alpha)=1$ if the trajectory initialized at $(r_c(0),\alpha(0))=(r_c,\alpha)$
converges to the spinning state and $\mathbf{1}_{\mathrm{spin}}(r_c,\alpha)=0$ otherwise. Here $r_{\max}$ is the maximum initial radius considered in the basin maps.

Figure~\ref{fig:basins}(d) shows $f_{\mathrm{spin}}$ (blue curve) and $r_c^{\mathrm{crit}}$ (red curve) as functions of the Stokes number. As in Fig.~\ref{fig:basins}(a--c), these quantities are computed using a uniform $200\times200$ grid of initial conditions in the $(r_c(0),\alpha(0))$ plane. For very small inertia, $f_{\mathrm{spin}}$ is nearly zero, indicating that almost no initial conditions reach the spinning state. This is consistent with the inertia-free limit, in which spirographic motion dominates for generic initial conditions, except for those initialized at the fixed points, $r_c(0)=0$ (spinning about the center-of-mass), and $r_c>0$ with $\alpha(0)=\pi/2$ (orbiting around the vortex center)~\cite{yerasi2022spirographic}. As the Stokes number is increased, $f_{\mathrm{spin}}$ becomes nonzero, reflecting the emergence of a finite basin of attraction for the spinning state. With further increase in $\mathrm{St}$, $f_{\mathrm{spin}}$ decreases again and tends to zero, consistent with the dominance of outward-spiraling trajectories at large inertia, as shown in Fig.~\ref{fig:basins}(c).

The critical radius $r_c^{\mathrm{crit}}$ decreases as $\mathrm{St}$ increases. Therefore, the maximum initial radius from which the spinning state can be reached increases as $\mathrm{St}$ decreases, although $f_{\mathrm{spin}}$ remains small at low $\mathrm{St}$. For sufficiently small $\mathrm{St}$, $r_c^{\mathrm{crit}}$ may exceed the sampled range $0 \le r_c(0) \le 2$. Nonetheless, within this sampled near-vortex range, the peak in $f_{\mathrm{spin}}$ at intermediate $\mathrm{St}$, together with the largest values of $r_c^{\mathrm{crit}}$, indicates that the spinning state is most likely to occur over an intermediate range of Stokes numbers.

\subsection{Linear Stability of the Spinning State}
Lastly, we examine the linear stability of the spinning state, defined by $\mathbf r_c=\mathbf 0$, $\mathbf v_c=\mathbf 0$, $\theta=\Omega_bt$, $\omega=\Omega_b$, where $\Omega_b=\Omega(r_b)$ and $r_b=\frac{1}{2}$ is the distance of each bead from the vortex center. By using $\bm\ell=(\cos\theta,\sin\theta)$, $\bm r_1=\bm r_c+\bm \ell/2$, $\bm r_2=\bm r_c-\bm\ell/2$, and $\bm u(x,y)=\Omega(x,y)(-y,x)$, the angular-velocity equation, Eq.~\eqref{eq:ndeqns-omega}, can be written as
\begin{equation} \label{eqn:omega_lsa}
\dot\omega=\frac{1}{2\mathrm{St}}\big[\mathcal F(\mathbf r_c,\theta)-\omega\big]    ,
\end{equation}
where 
\begin{equation}   
\mathcal F=\frac{\Omega_1+\Omega_2}{2}+(\Omega_1-\Omega_2)(x_c\cos\theta +y_c\sin\theta)
\end{equation}
with $\Omega_i=\Omega(r_i)$. When Eq.~\eqref{eqn:omega_lsa} is linearized about the spinning state, a perturbation of the center-of-mass position of magnitude $\delta\ll1$ implies $r_c = O(\delta)$. Expanding $r_{1,2} = |\mathbf r_c \pm \bm\ell/2|$ about $\mathbf r_c = 0$ gives $r_{1,2} = \tfrac12 \pm O(\delta)$, and hence $\Omega_1+\Omega_2 = 2\Omega_b + O(\delta^2)$. Likewise, $\Omega_1-\Omega_2 = O(\delta)$ and $x_c\cos\theta + y_c\sin\theta = O(\delta)$, so that the second term in $\mathcal F$ is $O(\delta^2)$. Thus, writing $\omega=\Omega_b+\delta\omega$, Eq.~\eqref{eqn:omega_lsa} reduces at linear order to
\begin{equation}
    \dot{\delta\omega}=-\delta\omega/2\mathrm{St},
\end{equation}
so the perturbation $\delta\omega$ decays exponentially and the rotational degrees of freedom are linearly stable. Moreover, the coupling between the center-of-mass motion and the rotational dynamics is $O(\delta^2)$ and can therefore be neglected at linear order. To study the stability of the spinning state, it is therefore sufficient to consider the center-of-mass dynamics. Since the spinning state is time-periodic, we formulate the perturbation dynamics in a frame co-rotating with the dumbbell, in which the spinning state becomes a fixed point.

We introduce the orthonormal co-rotating frame
\begin{equation}
\begin{aligned}
\mathbf e(t) &= (\cos(\Omega_b t),\,\sin(\Omega_b t)),\\
\mathbf e_\perp(t) &= (-\sin(\Omega_b t),\,\cos(\Omega_b t)).
\end{aligned}
\end{equation}
which satisfies $\dot{\mathbf e}=\Omega_b\,\mathbf e_\perp$ and $\dot{\mathbf e}_\perp=-\Omega_b\,\mathbf e$.

Decomposing perturbations in this frame as
\begin{equation}
\delta\mathbf r_c=\xi\,\mathbf e+\eta\,\mathbf e_\perp,\qquad
\delta\mathbf v_c=p\,\mathbf e+q\,\mathbf e_\perp,
\end{equation}
where $\xi$ and $\eta$ denote the center-of-mass position perturbations along
$\mathbf e$ and $\mathbf e_\perp$, respectively, and $p$ and $q$ are the corresponding
velocity perturbations.

Projecting the linearized equations of motion onto the co-rotating basis
$\{\mathbf e,\mathbf e_\perp\}$ yields an autonomous system for the perturbation variables.
In particular, $(\xi,\eta,p,q)$ satisfy
\begin{equation}
\dot{\xi}=p+\Omega_b\,\eta,\qquad
\dot{\eta}=q-\Omega_b\,\xi,
\label{eq:lin_kin}
\end{equation}
\begin{equation}
\begin{aligned}
\dot{p} &= \frac{1}{\mathrm{St}}\bigl(-\Omega_b\,\eta-p\bigr)+\Omega_b\,q,\\
\dot{q} &= \frac{1}{\mathrm{St}}\bigl(\bigl[\Omega_b+r_b\Omega_b'\bigr]\xi-q\bigr)
-\Omega_b\,p,
\end{aligned}
\label{eq:lin_dyn}
\end{equation}

where $\Omega_b'=\left.\dfrac{d\Omega}{dr}\right|_{r=r_b}$.

Equations~\eqref{eq:lin_kin}--\eqref{eq:lin_dyn} define a linear system
\(
\dot{\mathbf z}=\mathrm{J}_{\mathrm{rot}}\mathbf z
\)
for
\(
\mathbf z=(\xi,\eta,p,q)^{\mathsf T}
\),
with Jacobian
\begin{equation}
\mathrm{J}_{\mathrm{rot}}=
\begin{pmatrix}
0 & \Omega_b & 1 & 0\\
-\Omega_b & 0 & 0 & 1\\[6pt]
0 & -\dfrac{\Omega_b}{\mathrm{St}} & -\dfrac{1}{\mathrm{St}} & \Omega_b\\[10pt]
\dfrac{\Omega_b+r_b\Omega_b'}{\mathrm{St}} & 0 & -\Omega_b & -\dfrac{1}{\mathrm{St}}
\end{pmatrix}.
\end{equation}

The eigenvalues of $\mathrm{J}_{\mathrm{rot}}$ are the roots of the characteristic polynomial
\begin{equation}
\lambda^4+a_1\lambda^3+a_2\lambda^2+a_3\lambda+a_4=0,
\label{eq:charpoly_rot}
\end{equation}
where
\begin{equation}
\begin{aligned}
a_1 &= \frac{2}{\mathrm{St}}, 
\quad
a_2 = \frac{1}{\mathrm{St}^2}+2\Omega_b^2,\\
a_3 &= -\frac{2}{\mathrm{St}}\,\Omega_b
      \bigl(\Omega_b+r_b\Omega_b'\bigr),\\
a_4 &= \Omega_b^4.
\end{aligned}
\end{equation}
The Routh--Hurwitz criterion \cite{routh1877stability,seborg2016process} yields the following necessary and sufficient conditions for all
eigenvalues to satisfy $\Re(\lambda)<0$, where $\Re(\cdot)$ denotes the real part:
\begin{equation}
\begin{aligned}
a_1 &> 0,\qquad a_3 > 0,\qquad a_4 > 0,\\
a_1a_2 &> a_3, \qquad 
a_1a_2a_3 > a_1^2a_4+a_3^2.
\end{aligned}
\label{eq:RH_quartic}
\end{equation}

For $\mathrm{St}>0$ and $\Omega_b\neq 0$, the coefficients satisfy $a_1>0$ and $a_4>0$. For $a_3>0$, the inequality $a_1a_2>a_3$ follows from $a_1a_2a_3>a_1^2a_4+a_3^2$ and is therefore redundant. The conditions in Eq.~\eqref{eq:RH_quartic} then reduce to $a_3>0$ and $a_1a_2a_3>a_1^2a_4+a_3^2$. Introducing the local logarithmic slope of the angular velocity at the equilibrium radius,
\(
\chi\equiv -\frac{r_b\Omega_b'}{\Omega_b},
\)
these conditions may be written as
\begin{equation}
\chi>1,
\qquad
\frac{1}{\mathrm{St}^2\Omega_b^2}>
\ \frac{(\chi-2)^2}{\chi-1},
\label{eq:RH_compact}
\end{equation}
which provides an explicit local criterion for linear stability in terms of $\Omega_b$ and $\Omega_b'$.

The condition $\chi>1$ implies that $\Omega(r)$ decreases sufficiently rapidly in a neighborhood of $r=r_b$. Before specializing to the Lamb--Oseen vortex, we first consider the case in which the angular velocity is locally well approximated by a power law. If $\Omega(r)\sim r^{-n}$ near $r_b$, the logarithmic slope reduces to $\chi=n$. The stability condition in Eq.~\eqref{eq:RH_compact} then yields the critical Stokes number
\begin{equation}
\mathrm{St}_{\mathrm{crit}}(n)
=
\frac{1}{\Omega_b}\,\frac{\sqrt{n-1}}{|n-2|},
\qquad n>1,
\label{eq:Stcrit_general}
\end{equation}
with linear stability for $\mathrm{St}<\mathrm{St}_{\mathrm{crit}}(n)$ for $n\neq 2$ (or $\chi \neq 2)$.

\begin{figure}[t]
\centering
 \includegraphics[width=\linewidth]{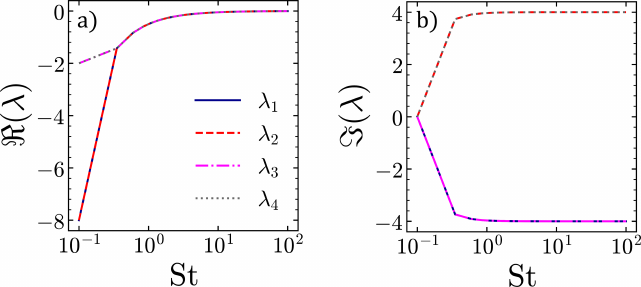}
\caption{ \label{fig:lastfig} Linear stability analysis of the spinning state. (a) Real parts of the four eigenvalues
$\lambda_{1,\dots,4}$ of $\mathrm{J}_{\mathrm{rot}}$ governing perturbations about the spinning state. (b) Imaginary parts of the same eigenvalues, showing the conjugate pairing $\pm\,\Im(\lambda)$.}
\end{figure}

For the Lamb--Oseen vortex, at $r_b/R = 5$, $\chi \approx 2$, a marginal case, for which the stability conditions Eq.~\eqref{eq:RH_compact} are satisfied for all finite $\mathrm{St}>0$. All roots of Eq.~\eqref{eq:charpoly_rot} therefore have a negative real part, and the spinning state is linearly stable. This is consistent with Fig.~\ref{fig:lastfig}(a), which shows that the real parts of all four eigenvalues, $\Re(\lambda)$ remain strictly negative. In the weak-inertia regime, the spectrum separates into two distinct decay rates, with one eigenvalue pair exhibiting substantially faster exponential relaxation than the other. As $\mathrm{St}$ increases, the real parts of all eigenvalues approach $0^{-}$, indicating a progressive weakening of the linear attraction toward the spinning state.

The imaginary parts of the four eigenvalues, $\Im(\lambda)$, form two complex-conjugate pairs (Fig.~\ref{fig:lastfig}(b)). For small $\mathrm{St}$, the imaginary parts are close to zero, corresponding to an essentially non-oscillatory (overdamped) relaxation. As $\mathrm{St}$ increases, the imaginary parts become finite and approach $\mathrm{St}$-independent limiting values. In this regime, the linear return to the spinning state is oscillatory, with a frequency that approaches a constant in the large-$\mathrm{St}$ limit.

Taken together, the basin-of-attraction structure and the linear stability analysis distinguish the local stability of the spinning state from its global attainability. The eigenvalue spectrum shows that the spinning state is linearly stable for all
$\mathrm{St}>0$. By contrast, the basin maps indicate that the set of initial conditions leading to the spinning state has finite measure only over an intermediate range of $\mathrm{St}$. The resulting non-monotonic dependence of $f_{\mathrm{spin}}$ and the reduction of $r_c^{\mathrm{crit}}$ at larger $\mathrm{St}$ therefore reflect not a loss of linear stability, but a loss of accessibility associated with the contraction of the basin of attraction.

\section{Conclusions and Perspectives}
\label{sec:Conclusions}

In this work, we examined inertial particle motion in a Lamb–Oseen vortex beyond the point-particle approximation. We considered a minimal model of a spatially extended particle: a rigid, symmetric dumbbell composed of two identical beads connected by a massless rod, and the hydrodynamic forcing is determined by the flow evaluated at the two bead positions.

As inertia is varied, the dumbbell exhibits three qualitatively distinct long-time behaviors. In the weak-inertia limit, the motion remains bounded and traces spirographic-like trajectories around the vortex center \cite{yerasi2022spirographic}. At sufficiently large inertia, outward drift dominates and the center-of-mass trajectories approach Fermat spirals, characteristic of inertial point-particle dynamics \cite{ravichandran2022waltz}. Between these limits, however, a distinct regime appears in which the dumbbell reaches a spinning state, with its center-of-mass trapped at the vortex center while the dumbbell rotates steadily.

Basin-of-attraction maps show that the structure of initial conditions leading to the spinning state changes qualitatively with inertia. At low Stokes numbers, trapping depends sensitively on both the initial radius and orientation and appears in a nested, banded pattern in the space of initial conditions. At intermediate inertia, this banded structure largely disappears, leaving a dominant connected region of initial conditions leading to spinning, together with a near-core trapping region. At larger inertia, trapping becomes confined to a progressively smaller region close to the vortex center. Aggregate measures quantify these trends: the fraction of initial conditions leading to trapping varies non-monotonically with the Stokes number, attaining finite values primarily at intermediate inertia and decreasing again at larger inertia, as reflected by the reduction of the critical initial radius.

The linear stability analysis shows that the spinning state is locally stable when the logarithmic slope of the angular velocity at the equilibrium radius exceeds a critical value. For vortices whose angular velocity decays as a power law, the slope is constant, and the stability criterion yields a critical Stokes number above which the spinning state becomes linearly unstable. The Lamb–Oseen vortex corresponds to a marginal case, for which the spinning state remains linearly stable for all Stokes numbers. The relaxation dynamics, however, depend on inertia, transitioning from overdamped decay at small inertia to damped oscillatory return at larger inertia, with progressively weaker attraction as inertia increases.

The mechanisms identified here show how nonlocal flow sampling, combined with inertia, can qualitatively alter particle-vortex interactions and induce long-time behavior distinct from point-particle dynamics. Background strain or time-dependent vortical flows provide a setting in which the persistence of the spinning state beyond a steady vortex can be assessed. Relaxing symmetry through unequal beads, either in mass or size (drag coefficient), would clarify the role of geometric and hydrodynamic balance. Adding bead-bead hydrodynamic interactions, Brownian forcing, or elastic connectors would further probe how the spinning state and its basin structure respond to additional couplings.

\appendix

\section{}
\label{sec:AppendixB}
\begin{figure}[!htb]
\centering
 \includegraphics[width=0.5\linewidth]{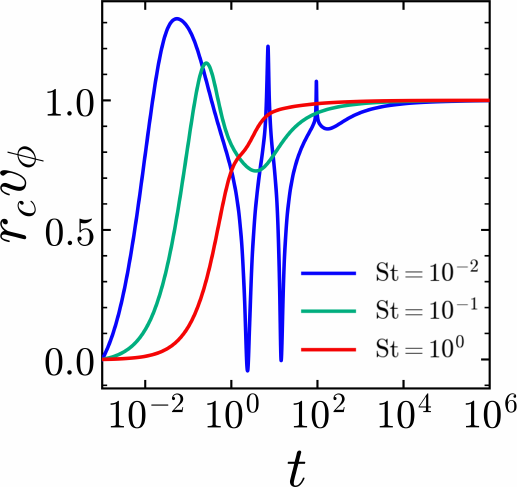}
  \captionsetup{justification=raggedright}
\caption{\label{fig:appendixB} Time evolution of $r_c v_\phi$ for outward-spiraling trajectories at different $\mathrm{St}$.}
\end{figure}

\begin{acknowledgments}
All authors acknowledge R.~Govindarajan (ICTS-TIFR) for helpful discussions. K.C. acknowledges support from the Indian Institute of Technology Indore under the Young Faculty Research Seed Grant Scheme. K.C. thanks E. D. Mackay (SLS-UoD) for helpful discussions. S.R.Y acknowledges the financial support from the CNRS through the 80 $|$ Prime program. S.R. is supported through IIT Bombay Seed Grant RD/0522-IRCCSH0-020. S.K. acknowledges the use of PARAM UTKARSH HPC facility of the National Supercomputing Mission project and thank center for Cyber Physical Systems (CCPS), National Institute of Technology Karnataka, Surathkal for the financial support for the same. D.V. acknowledges the support of Agence Nationale de la Recherche through Project No. ANR-21-CE30-0040-01 and the Indo–French
Centre for the Promotion of Advanced Scientific Research \linebreak (IFCPAR/CEFIPRA, Project No. 6704-1).
\end{acknowledgments}


%

\end{document}